\documentstyle[twocolumn,psfig,floats,aps]{revtex}

\begin{document}
\twocolumn[\hsize\textwidth\columnwidth\hsize\csname
@twocolumnfalse\endcsname
\title{\bf Low-frequency current fluctuations in doped ladders}
\author{Kenji~Tsutsui$^a$, Didier~Poilblanc$^{b,a}$ and
  Sylvain~Capponi$^a$}
\address{
$^a$Laboratoire de Physique Quantique \& UMR--CNRS 5626,
Universit\'e Paul Sabatier, F-31062 Toulouse, France \\
$^b$Theoretische Physik, ETH-H\"onggerberg, CH-8093 Z\"urich,
Switzerland
}
\date{\today}
\maketitle

\begin{abstract}
Charge current static and dynamical correlations are computed by exact
diagonalisation methods on a 2-leg t-t'-J ladder which exhibits a
sharp transition between a Luther-Emery (LE) phase of hole pairs 
and a phase with deconfined holes.
In the LE phase, we find short-range low-energy
incommensurate current fluctuations which are intrinsically connected
to the internal charge dynamics within one hole pair. 
On the contrary, when holes unbind, the maximum of the
current susceptibility moves abruptly to 
the commensurate wavevector $\pi$ and strongly increases for decreasing
doping suggesting an instability towards a staggered flux  state
at sufficiently small doping.

\smallskip

\end{abstract}

\vskip2pc]



Among mean-field solutions for two-dimensional (2D)
strongly correlated electrons,
the staggered flux phase~\cite{Affleck_Marston} appeared as 
an appealing candidate. Due to the
local SU(2) symmetry, the staggered flux  phase is
equivalent, at half-filling, to the projected d-wave pairing BCS 
variational wavefunction~\cite{Gros}. Both states could equivalently be 
viewed as a Resonating Valence Bond state~\cite{RVB}. 
Since hole doping lifts the degeneracy between the staggered flux  state
and the d-wave pairing state, it is of importance to investigate 
possible competition between such states. 
At finite doping, as the newly proposed d-density wave state\cite{Bob},
the staggered flux  state would break translation as well as time reversal
symmetry. In otherwords, it would exhibit "orbital antiferromagnetism"
i.e. a checkerboard pattern of plaquette currents circulating
clockwise and anti-clockwise on the two different sublattices.
On the other hand, it has been shown~\cite{PALee} recently 
that the current vorticity correlations in a variational d-wave
pairing state follows closely
the hole correlations within one pair.

Numerical investigations carried out on small (2D) clusters have shown that
staggered or commensurate (with a larger 
period modulation) flux states are very good variational
states~\cite{num_var}. In addition, spin-chirality correlations have
been tested numerically in the 2D t-J model~\cite{num_exact} 
showing that staggered (i.e. at ${\bf q}=(\pi,\pi)$)
chiral fluctuations are strongly enhanced with increasing hole kinetic energy.
Recent calculations of the static current-current correlations on a 32-site
lattice with two holes~\cite{Leung} have confirmed the existence of 
staggered vorticity in this model. 

In this paper, search for current fluctuations have been carried out 
in an extended t-J model on a two-leg ladder where finite size
effects can be handled in a controled fashion.
Such a system is known
to exhibit a robust spin gap at and close to half-filling as well as
hole pair binding~\cite{DR96,HPNSH95}. Dominant power-law d$_{x^2-y^2}$-like
pairing and $4k_F$ charge density wave (CDW) correlations
at small doping are characteristic of a Luther-Emery (LE) liquid 
regime~\cite{LE96}. 
An additional {\it negative} diagonal hopping~\cite{cuprates} $t'$ between
next-nearest neighbor (NNN) sites was shown to
strongly reduce hole pair binding
leading e.g. to the emergence of low-energy quasiparticle-like triplet 
excitations~\cite{tprime_ladder}. On the contrary, a positive $t'$
increases the pair binding energy even further. 
Here, we use this parameter to monitor the hole pair binding energy in
order to investigate possible connections between staggered-vorticity
correlations and hole correlations. We find that incommensurate 
current fluctuations occur on the scale of the pair size. 
For sufficiently negative $t'$ hole pairs break
appart~\cite{tprime_ladder} and low energy staggered vorticity 
fluctuations are found as well as static commensurate correlations
in the ground state (GS). We argue that these finding might signal
an instability towards a staggered flux state. 

The two-leg ladder Hamiltonian reads,
\begin{eqnarray}
H&=&  J_{\rm rung} \sum_{i} ({\bf
  S}_{i,1} \cdot {\bf S}_{i,2}-\frac{1}{4}n_{i,1}
n_{i,2}) \nonumber \\
&+& J_{\rm leg}\sum_{i,l} ({\bf
  S}_{i,l} \cdot {\bf S}_{i+1,l}-\frac{1}{4}n_{i,l}
n_{i+1,l})\\
&+& \sum_{i,j,l,m,\sigma} t_{ij}^{lm}
(c_{i,l,\sigma}^\dagger c_{j,m,\sigma} + h.c.) \nonumber 
\label{Ham}
\end{eqnarray}
where $i$ is a site index along the chain direction, $l$ and 
$m$ (=1,2) label the 2 legs and $n_{i,l}$ is the local
charge density operator. Holes are assumed to hop along the
chain direction or between legs with amplitudes 
$t_{i\, i\!+\!1}^{l\, l}=t_{\rm leg}$ and 
$t_{i\,  i}^{1\, 2}=t_{i\,  i}^{2\, 1}=t_{\rm rung}$ respectively. In
addition, as stated above, it is of particular interest 
to include a NNN hopping 
$t_{i\, i\!\pm\!1}^{1\, 2}=t_{i\, i\!\pm\!1}^{2\,1}=t^\prime$ \cite{cuprates}.
In the following, for sake of simplicity,
we shall only discuss the case of a spatially isotropic
ladder i.e. with $t_{\rm leg}=t_{\rm rung}=t$ (set to 1) and 
$J_{\rm leg}=J_{\rm rung}=J$. In addition, $J$ is set to 0.5.
However, we believe our results are generic.
The static and dynamical quantities reported in this paper
have been obtained by exact diagonalisations (ED) of finite periodic 
2$\times$L clusters (L=8 to L=12) doped with up to 6 holes.

\begin{figure}[htb]
\vspace{-0.2truecm} 
\begin{center}
\psfig{figure=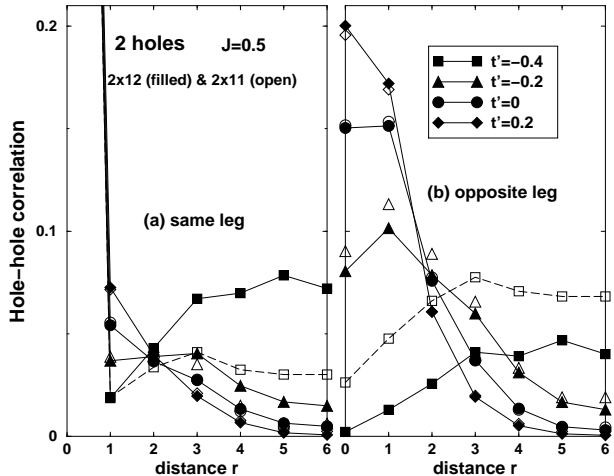,width=8.0truecm,angle=-90}
\end{center}
\caption{Hole-hole (equal-time) correlation function on a two 
hole-doped 2$\times$12 ladder vs separation between holes
in the same leg (a) and
in different legs (b).
Different filled symbols correspond to different $t'$ parameters
according to the legend.
Data obtained on a 2$\times$11 ladder are also shown as open symbols (unless
undistinguishable from the previous sets). Data are normalized to the 
hole density i.e. such that the local ($r=0$) correlation equals 1.}
\label{hole_corr}
\end{figure}

The NNN hopping $t'$ is the key parameter of our study since it 
monitors the attraction between holes in the system~\cite{tprime_ladder}.
To demonstrate this fact, as a preliminary study, we have computed the
(static) charge correlations on 2 hole-doped ladders for many different
parameters $t'$. Selections of these data reported in
Fig.~\ref{hole_corr} show clearly 2 different behaviors; above a 
critical value $t_c^\prime$ (we find $t_c^\prime\simeq -0.3$ at $J=0.5$) the
two holes are bound and the hole-hole correlation decreases very fast
with distance. Note that, in this case, finite size effects remain
quite small since hole correlations at short distance are rather 
independent on system size. 
The bound pair "size" increases with decreasing $t'$
and eventually at $t'=t_c^\prime$ (negative) the GS
changes abruptly\cite{note_critical} into a state where holes are unbound.
Indeed, as shown for $t'=-0.4$ in Fig.~\ref{hole_corr}, 
the hole-hole correlation function {\it increases} with distance 
which means that holes tend to avoid  each other.
It is interesting to notice that boundary effects become quite
important in this case probably in connection with the
enhancement of the antiferromagnetic (AF) correlations.
In order to avoid magnetic frustration, the 2 holes tend to 
be located, on average, on the same (opposite) leg for even (odd)
ladder lengths~\cite{note_parity}. In the following, we shall show 
that the charge
current fluctuations also change rapidly at the transition between 
bound pairs and individual holes.

The current operators for the rung and leg bonds are defined as
follows. For the leg,
\begin{equation}
j_q  =  - it_{{\rm leg}} \sum\limits_{j,l,\sigma } {e^{i\left( {qj + \pi l} 
\right)} \left( {c_{j + 1,l,\sigma }^ +  c_{j,l,\sigma }^{}  
- h.c.} \right)} \, ,
\end{equation}
where $l$ and $n$ denote the site numbers for the leg and rung
directions, respectively. Note that the leg current is defined in such
a way that current densities on opposite bonds on the two legs have
opposite signs as it should for {\it loop} currents defined on the
plaquettes. For the rung,
\begin{equation}
j_q  =  - it_{{\rm rung}} \sum\limits_{j,l,\sigma } {e^{iqj} \left(
    {c_{j,l + 1,\sigma }^ +  c_{j,l,\sigma }^{}  - h.c.} \right)} \, .
\end{equation}
In the staggered flux  phase~\cite{Affleck_Marston} both rung and leg 
currents would acquire a finite expectation value at momentum $q=\pi$ 
corresponding to a staggered arrangement of the loop currents on the 
plaquettes. Here we consider a more general situation of possible
{\it incommensurate} modulations with arbitrary q~\cite{note_incom} 
of the current pattern along the leg direction. Dynamical fluctuations of 
such loop currents can be investigated by computing the
associated complex dynamical susceptibility,
\begin{equation}
\chi_{jj} \left( {q,\omega } \right) \equiv \sum\limits_n 
\frac{\left| \left\langle n \right|j_q\left| 0 \right\rangle 
\right|^2 }{\omega+i\epsilon +E_n  - E_0 }  \, ,
\end{equation}
where $\left| n \right\rangle$ ($E_n$) are exact eigenstates
(eigenenergies) of the hamiltonian.
It is instructive to define the current-current dynamical correlation 
function $S_{jj}$ as the imaginary
part of the susceptibility devided by frequency,
\begin{eqnarray}
S_{jj}\left(q,\omega\right)&\equiv&
\frac{\chi_{jj}^"\left( {q,\omega } \right)}{\omega}
\nonumber \\
&=& \sum\limits_n {\frac{{\left| {\left\langle n \right|j_q
\left| 0 \right\rangle } \right|^2 }}{{E_n  - E_0 }}\delta \left( 
{\omega  - E_n  + E_0 } \right)} \, ,
\end{eqnarray}
which leads, after integration over frequency, to the static
($\omega=0$) current susceptibility (Kramers-Kr{\"o}nig),
\begin{equation}
\chi_{jj} \left( q,0 \right) 
= \int {\frac{d\omega}{\omega} \chi _{jj}^" \left( {q,\omega } \right)}  
= \sum\limits_n {\frac{{\left| {\left\langle n \right|
j_q \left| 0 \right\rangle } \right|^2 }}{{E_n  - E_0 }}} \, .
\end{equation}
The current-current dynamical correlation $S_{jj}$ and the related
static susceptibility have been computed on finite clusters by a
standard continued-fraction method. Note that the susceptibility
$\chi(q,0)$ can alternatively be calculated as a response to a
{\it fictitious} modulated magnetic flux. We have checked that this 
second method gives identical results. We also have found
that, quite generally, use of bond and leg currents gives very similar
results (especially at $q=\pi$) 
so that in the following, for sake of clarity, 
only results using the {\it rung} current operator will be presented.

\begin{figure}[htb]
\vspace{-0.2truecm} 
\begin{center}
\psfig{figure=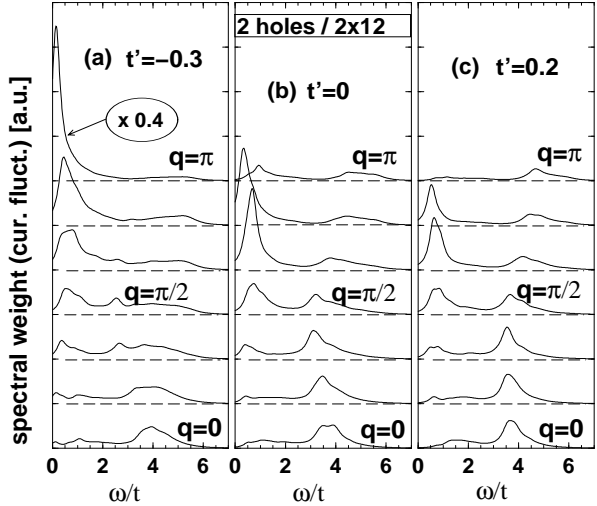,width=8.0truecm,angle=0}
\end{center}
\caption{Dynamical correlation function $S_{jj}(q,\omega)$ calculated
on a 2 hole-doped periodic ladder for $J=0.5$. Each panel corresponds to
a given value of $t'$ as indicated. For clarity, curves obtained for 
all available momenta $q$ are shifted upwards w.r.t. each other by an
equal amount. From bottom to top, $q=2\pi n/L$, $n$ from 0 to $L/2$.
Note the reduction factor of 0.4 applied to the data at 
$t'=-0.3$ and $q=\pi$.
}
\label{curr_dyn1}
\end{figure}

\begin{figure}[htb]
\vspace{-0.2truecm} 
\begin{center}
\psfig{figure=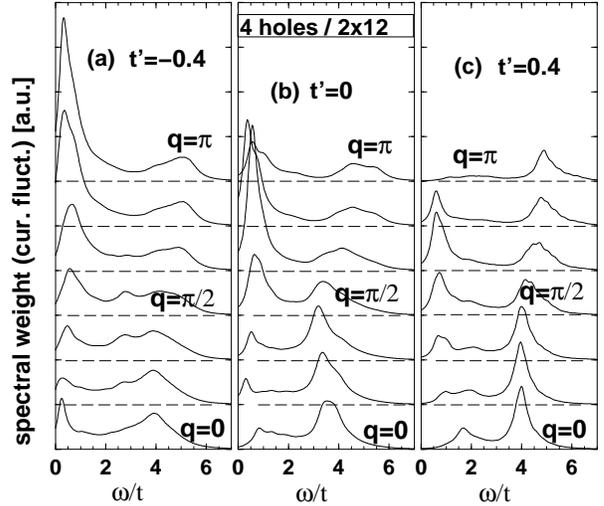,width=8.0truecm,angle=0}
\end{center}
\caption{Dynamical correlation function $S_{jj}(q,\omega)$ calculated
on a 4 hole-doped periodic ladder for $J=0.5$. 
Same presentation as Fig.~\protect\ref{curr_dyn1}
}
\label{curr_dyn2}
\end{figure}

We first focus on a ladder doped with only two holes. A selection
of our results obtained on a 2$\times$12 ladder are presented 
on Fig.~\ref{curr_dyn1}. On should distinguish two interesting
features on these data; (a) some high energy structure in the range
$\omega\sim 4t$ which is merely dispersing and whose weight
increases with increasing $t'$; (b) a low energy structure at 
a fraction of J. Above $t_c^\prime\simeq -0.3$, the maximum 
weight of this structure is spread on incommensurate~\cite{note_incom} 
wavevectors in the vicinity of $\pi$. On the contrary, when the
two holes become unbound ($t'<t_c^\prime$), the largest low energy 
(typically $\omega_{\rm typ}\sim 0.2J=0.1t$) current fluctuations 
occur at $q=\pi$ corresponding to
the staggered current pattern of the staggered flux  state.
The same behavior is also observed for higher doping as can be 
seen in Fig.~\ref{curr_dyn2} showing similar data but for a doping
of four holes. 

\begin{figure}[htb]
\vspace{-0.2truecm} 
\begin{center}
\psfig{figure=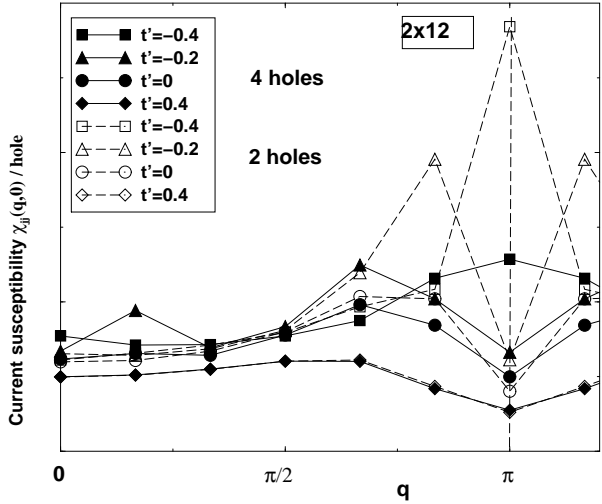,width=8.0truecm,angle=-90}
\end{center}
\caption{Current-current susceptibility vs momentum q for
different NNN hopping $t'$ (as indicated on the plots)
calculated on 2 and 4-hole doped 2$\times$12 ladders.
The susceptibilities have been normalised w.r.t. the  
number of holes to emphasize proportionality if any. Arbitrary
units are used. Note that at $t'=0.4$ data for 2 and 4 holes are
almost indistinguishable. 
}
\label{curr_suscep}
\end{figure}

The behavior of the integrated spectral weight, namely the
current-current susceptibility, as a function of parameters and doping
is also of particular interest. Data reported in
Fig.~\ref{curr_suscep} indeed reveal that a qualitative change
also occurs at the transition between bound hole pairs and free holes.
When holes are tighly bound into pairs (typically for $t'>0$) the
current susceptibility is rather featureless and {\it scales with the 
number of holes}. When $t'\rightarrow t_c^\prime$ from above, 
broad peaks appear in the vicinity of momentum $\pi$~\cite{Nofq}
especially in the system containing only two holes. 
This increase of total spectral weight is mainly due to the increase
of the low energy incommensurate peaks in the dynamical correlation
functions shown in Figs.~\ref{curr_dyn1}(b),~\ref{curr_dyn2}(b).
When hole pairs break apart i.e. for $t'<t_c^\prime$ a large peak 
appears at the commensurate wavevector $q=\pi$.

\begin{figure}[htb]
\vspace{-0.2truecm} 
\begin{center}
\psfig{figure=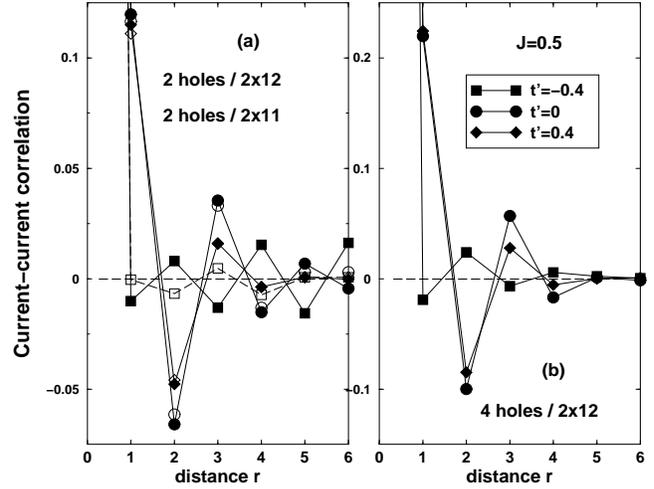,width=8.0truecm,angle=-90}
\end{center}
\caption{Current (equal-time) correlation function vs distance for two (a)
and four (b) hole-doped ladders calculated on a 2$\times$12 cluster.
Different filled symbols correspond to different $t'$ parameters
according to the legend.
In (a) data for a 2$\times$11 ladder are also shown as open symbols (unless
undistinguishable from the previous sets).}
\label{curr_corr}
\end{figure}

To complete our study, we have also computed the {\it equal-time}
current-current correlation function 
$\sum_q e^{iqr}\langle j_q j_{-q}\rangle$ vs separation r between 
two rungs. Recent Density Matrix Renormalisation Group (DMRG) studies
on a 2 leg t-J ladder have shown that such correlations are 
exponentially suppressed with distance 
in the LE liquid phase~\cite{Doug}. Our data for various $t'$ and 2
different hole concentrations are plotted in Fig.~\ref{curr_corr}.
Above $t_c^\prime$, in the LE liquid phase, the alternating current
correlations are indeed rapidly decreasing with distance
in agreement with the DMRG results. When holes
are strongly bound into pairs (as it is the case e.g. for $t'=0.4$)
it is interesting to notice that,
apart from an overal scaling factor proportional to the hole
concentration, the current correlation is almost completely
independent of the number of pairs in the system. This is a further
confirmation that current fluctuations occur within one pair.
On the contrary, when pairs break apart the current correlations within
the GS increase strongly~\cite{tJV_current} with decreasing hole density.
In this case, establishment of staggered current correlations depends
strongly on the parity of the ladder length because of possible
frustration as seen in Fig.~\ref{curr_corr}(a).

We finish this paper by briefly discussing the role of the SU(2)
symmetry of the model. As mentioned above, it leads to the
equivalence between d-wave pairing and staggered flux  phases at
half-filling~\cite{Gros}. Since this degeneracy 
is lifted upon doping, the hole 
kinetic energy is expected to stabilize one phase over the other.
The sudden change of properties in a t-t'-J ladder at small doping
when varying $t'$ is consistent with this scenario if one
assumes that the competition between the pairing and the staggered flux 
phases is indeed governed by the small kinetic energy term
(proportional to doping) and hence by $t'$. 

In summary, our data strongly suggest that, when hole pairs are stable, the 
current fluctuations is intrinsically related to the internal
spin/charge dynamics within one pair at least in the regime where
pairs do not overlap. The magnitude of the current fluctuations
increases as the spatial extension of the pairs increases
until they start to overlap. These findings are consistent 
with the picture of bound 
hole pairs of holes carrying opposite current vorticity~\cite{PALee}.
When hole pairs break apart into individual
holes, commensurate staggered fluctuations set in. The increase
of the magnitude of these fluctuations as one moves towards vanishing
doping is interpreted as an instability to a
(static) staggered flux  state.

Computations were performed on the NEC-SX5 at IDRIS, Orsay (France). 
D.P. acknowledges support from the Center for Theoretical Studies 
and the Institute for Theoretical
Physics at ETH-Z\"urich and thanks T.M.~Rice for useful discussions.


\begin{thebibliography}{99}
\bibitem{note_critical} Note that in most cases (e.g. 2 holes on
2$\times$10 and 2$\times$12 lattices, 4 holes on 2$\times$12, etc...) 
this transition is linked to a level crossing between two GS
wavefunctions with different total momenta.
\bibitem{note_incom} By "incommensurate" we, in fact, mean
commensurate modulations with larger periodicities.
\bibitem{Nofq} Note that the static charge structure factor 
exhibits a sharp "4k$_F$" singularity at $q=2\pi n$ for $t'\ge 0$
as expected in a LE liquid. This feature tends to quickly deasappear for
increasingly negative $t'$. Details will be reported elsewhere.
\end{thebibliography}
\end{document}